# What We *Don't* Know About Spreadsheet Errors Today:

The Facts, Why We Don't Believe Them, and What We Need to Do

Raymond R. Panko
Shidler College of Business
University of Hawai`i
2404 Maile Way
Honolulu, HI 96822
ray@panko.com

**ABSTRACT**

Research on spreadsheet errors is substantial, compelling, and unanimous. It has three simple conclusions. The first is that spreadsheet errors are rare on a per-cell basis, but in large programs, at least one incorrect bottom-line value is very likely to be present. The second is that errors are extremely difficult to detect and correct. The third is that spreadsheet developers and corporations are highly overconfident in the accuracy of their spreadsheets. The disconnect between the first two conclusions and the third appears to be due to the way human cognition works. Most importantly, we are aware of very few of the errors we make. In addition, while we are proudly aware of errors that we fix, we have no idea of how many remain, but like Little Jack Horner we are impressed with our ability to ferret out errors. This paper reviews human cognition processes and shows first that humans cannot be error free no matter how hard they try, and second that our intuition about errors and how we can reduce them is based on appallingly bad knowledge. This paper argues that we should reject any prescription for reducing errors that has not been rigorously proven safe and effective. This paper also argues that our biggest need, based on empirical data, is to do massively more testing than we do now. It suggests that the code inspection methodology developed in software development is likely to apply very well to spreadsheet inspection.

## 1 INTRODUCTION

Research on spreadsheet errors is substantial, compelling, and unanimous. As we will see later, all empirical studies, without exception, have found errors at frequencies that few would call acceptable. Actually, if we had done no spreadsheet research at all, results from human error research in other fields would have given us all the data we need to treat spreadsheet error risks seriously.

Yet few developers or corporations have responded to the error research by the only means that has proven effective in other human cognitive domains: massive testing. The problem seems to be that the error rate results do not "feel right". In our experience, we do not make as many errors as the literature says we do, we catch far more errors than the literature says we do, our spreadsheets are rarely materially incorrect, and the steps we take to make our spreadsheets error free are effective. In the words of American comedian Stephen Colbert, the results that suggest otherwise lack the feeling of "truthiness." This is true despite the fact that every empirical study has found overconfidence in the face of many errors.

At the root of this disconnect between research findings and our error beliefs, this paper argues, are the fundamental ways in which human attention and cognition work. We will begin the discussion by looking at how our cognition presents the world to us. We will see that our cognition edits out anything that is unimportant, including most errors. This distorts our "experience" of errors, and this distorted experience is what drives our intuition about error risks. Furthermore, our cognition makes us confident that our "constructed reality" is accurate. Pilots learn that when they are flying through a cloud, they need to trust the data from their instruments, not their perceptions. Business professionals arguably need to do the same when they deal with how they should build and test spreadsheets.

If we do accept spreadsheet error research as being correct, then what should we do in response? We will argue that serious error reduction is likely to be predominantly about testing but not testing as we now do it. Testing based on research will be far more expensive. The "bottom line" is that spreadsheet programs are not error-prone. People are error prone. To use spreadsheet programs is not wrong, but using those that are not tested by research-based methods is.

## 2 THE EVIDENCE

We will not go over the evidence for the prevalence of spreadsheet errors in detail. Tables of results can be found at panko.com, under both human error and spreadsheet error research. Spreadsheet error research is a tiny corner of the human error research field, so much research on spreadsheet errors has focused on seeing if human error research results apply to spreadsheet error. The brief answer is that they apply very well.

### 2.1 Error Rates during Development

In the 1980s, there was a kind of Grand Unification in human error research. Researchers in different domains found that their error rates were nearly identical for tasks of comparable cognitive complexity. These error rates were reported as base error rates (BERs), which are long-term average error rates quantified as the number of errors per 100 actions.

BERs obviously depend on complexity. For simple but nontrivial cognitive actions such as writing, calculating, and writing program statements, BERs are usually in the range of 1% to 5%. This is found in both experiments and data collection in real organizations. Programming BERs are based on particularly large corpuses of actual software testing in corporations. Table 1 shows four such corpuses totaling around 10,000 code inspections in industry. These data come from the Panko human error website, which also has data on smaller corpuses. The average error rates of the four corpuses ranged from 1.9% to 3.7%.

**Table 1: Error Rates in Program Statements and Spreadsheet Cells**

| | | |
|---|---|---|
| More than 6,000 code inspections in industry, per line of code. | 1.9% | Weller [1993] |
| National Software Quality Experiment, per line of code. | 2.0% | O'Neill [1994] |
| 2,500 inspections at Cisco Systems, per line of code. | 3.2% | Cohen [2006] |
| AT&T. 2.5 million lines of code over 8 software releases, per line of code. | 3.7% | Graden and Horsley [1986] |
| 14 laboratory studies of spreadsheet development, 967 participants working alone on a variety of tasks. Cell error rate | 3.9% | Panko Spreadsheet Research Website |

Based on human error research, we would expect spreadsheet cell error rates to in the range of 1% to 5%. Table 1 summarizes results from 14 laboratories spreadsheet development studies involving 967 individuals working alone on a variety of tasks. These numbers come from the Panko spreadsheet research website. The average across these studies is a cell error rate of 3.9%.

These per-cell error rates are small. However, spreadsheets contain many cells, and the probability of an error increases rapidly when there are many calculations that depend on precedent cells. Figure 1 shows the importance of error compounding over chains of calculations. The top line in the figure is for a spreadsheet with 100 cells in cascades. This might be a single cascade of 100 cells, 25 cascades of 4 cells, 10 cascades of 10 cells, or any other combination. This might be 100 root (non-copied) formulas, although copying is not error-free, especially when the formula changes in "copied" cells in complex ways.

**Figure 1: Cell Error Rates and Probabilities of a Bottom-Line Error**

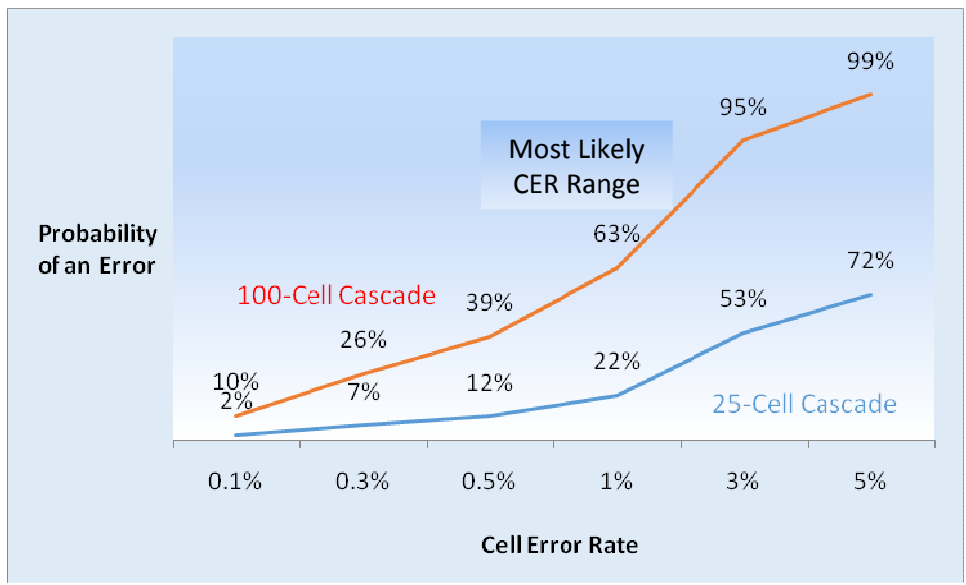

Spreadsheets with 100 or more cells in cascades are plausible to consider given the size of corporate spreadsheets. Hermans and Murphy-Hill [2015] analyzed spreadsheets obtained during the legal discovery process at Enron. This corpus included 9,120 Enron spreadsheets with formulas. In these spreadsheets, they found 20,277,835 cells with formulas. Other censuses of spreadsheets have also revealed massive numbers of large spreadsheets. For example, McDaid, et al. [2011] found 65,806 spreadsheets on the department servers in two organizations. These spreadsheets averaged over 4,000 formulas. Even with extensive copying, the number of root formulas must have been extremely large in both corpuses.

Figure 1 shows that for cell error rates of 1% to 5%, the probability of a calculation error is overwhelming. In fact, even error rates an order of magnitude lower would give error probabilities that any company would view as unacceptable. A second line gives the probability of a bottom line error in a spreadsheet with a mere 25 cells in cascades. Here too, likely cell error rates give lower but still unacceptable error rates.

Are bottom-line error rates in real spreadsheets consistent with Figure 1? Table 2 shows that the answer is *yes*. The table shows the results of 85 intensive inspection studies that took several days per spreadsheet. Most were done by commercial spreadsheet "auditing" firms. One was done by a government tax collection unit using a strong methodology that included substantial cell-by-cell inspection. Hicks [1991] used software code inspection methodology with a team of three inspectors to examine a spreadsheet module. Collectively, these studies found errors in 94% of the spreadsheets studied.

**Table 2: Errors Found in Operation Spreadsheets during Intensive Inspection**

| | | | | |
|---|---|---|---|---|
| Hicks [1995] | 1 | 100% | 1.2% | Fagan Code Inspection with a Team of Three |
| Coopers and Lybrand [1997] | 23 | 91% | | Commercial "Audit" |
| KPMG [1998] | 22 | 91% | | Commercial "Audit" |
| Lukasic [1998] | 2 | 100% | 2.2%, 2.5% | Reproduction in a Financial Modeling Language |
| Butler [2000] | 7 | 86% | | Audit of Spreadsheet Submitted with Taxes in the United Kingdom |
| Lawrence and Lee [2001] | 30 | 100% | | Commercial "Audit" |
| Total / Weighted Average | 85 | 94% | | |

Note also that two studies reported cell error rates. These are similar to those expected from human error research and similar to those found in spreadsheet development experiments.

## 2.2 Errors Detected During Inspection

Given likely error rates in operational spreadsheets, extensive testing will be needed—more extensive than the testing that was done on operational spreadsheets.

Effective testing is likely to be difficult and extremely expensive. Although people are 95% to 99% accurate when they do calculations, write, code, or enter spreadsheet cells, human error research has shown that humans are much worse at finding errors that have occurred.

Data from the Panko human error website on proofreading for spelling shows that detection rates are 81% for simple spelling errors but only 66% for more complex spelling errors. Detection rates plummet for longer words [Healey, 1980] and more complex material [Riefer,1991].

In software development, error detection rates are only 20% to 40% [Basili and Selby, 1986; Johnson and Tjahjono, 1997; Myers, 1978; Porter, Votta, and Basili, 1995; Porter and Johnson, 1997; Porter, Sly, Toman, and Votta, 1997; Porter and Votta, 1994] when single inspectors examine a code module to look for errors. This is why software code inspection is always done in teams of three to five or more [Fagan, 1976 1986; Gilb and Graham, 1993; Cohen, 2006].

In spreadsheet inspection, nine experiments using over 1,000 participants in total have found average error detection rates of 60%. However, detection rates depend heavily on error complexity. Even simple and obvious errors are not detected with close to 100 percent accuracy. Anderson [2004] published data for different types of error in his study, which was done by experienced spreadsheet developers. Of 14 errors with different ranges of complexity, three had detection rates below 10%, and half had detection rates below 40%. Even the simplest errors were caught by fewer than 60% of the participants, and no participant found all errors.

What type of errors do people make when they build spreadsheets? Panko and Halverson [1996] divided spreadsheet errors into logical (mathematical and domain knowledge) errors, mechanical errors (such as pointing to the wrong cell and typing errors), and omission errors (leaving out a necessary component). In one experiment, Panko and Halverson [2001] Found that 45% of errors that participants made were logic errors, 23% were mechanical errors, and 31% were omission errors. In another study using a different task, Panko and Sprague [1998] found that 44% of the total errors were logic errors, 3% were mechanical errors, and a whopping 53% were omission errors. The large number of omission errors in this study was an anomaly of the task statement wording. The statement specified that "Both walls will be built by crews of two. Crews will work three eight-hour days to build either type of wall." Almost all omission errors occurred when the participants omitted the number of members per crew, the number of days per wall, or the number of hours per day. Their memory was overloaded. This type of error is called a lapse in human error research. In an inspection of an operational spreadsheet using the Fagan [1976, 1986] methodology, Hicks [1995] found that most errors in the spreadsheet his team examined were logic errors. Omission errors were also widely seen in spreadsheet studies. In the Brown and Gould [1987] experiment, when omission errors were excluded, 44% of the spreadsheet models had errors. When omission errors were included, this rose to 63%. Omission errors are especially dangerous because their detection rate tends to be very low, especially if the omission error is due to a misinterpretation of the situation [Reason, 1990; Woods 1984].

## 2.3 Inexperience, Rushing, and Exacerbaters

Rushing, inexperience, code complexity, stress, and other factors can exacerbate error rates. However, as Figure 2 illustrates, reducing exacerbaters does not eliminate errors. There usually is a floor beyond which error reduction is minimal. Under extremely unchallenging conditions, error rates may even increase.

**Figure 2: Error Exacerbation**

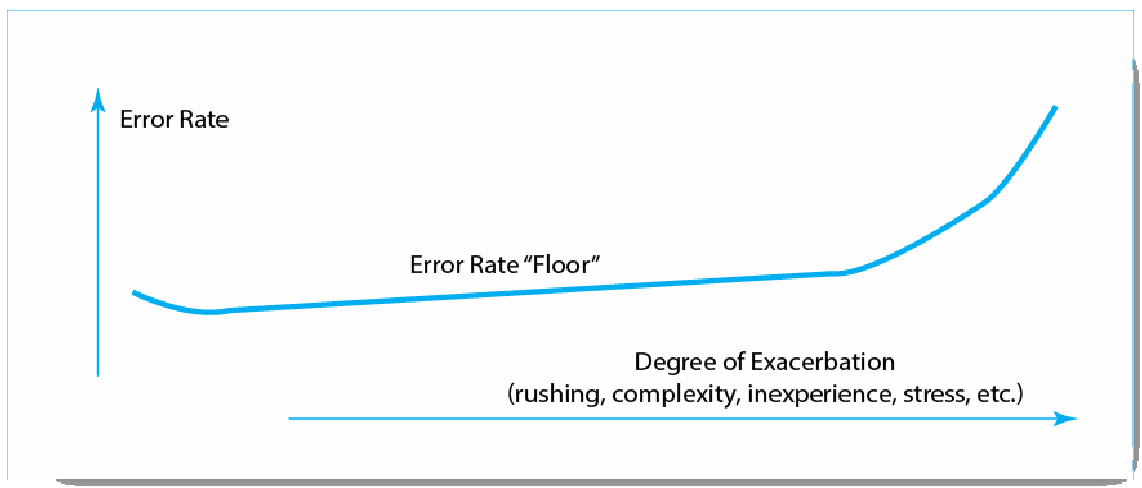

Table 3 shows how errors relate to stress [Swain and Guttman, 1983]. It shows multipliers of "normal" BERs under different conditions. If the BER under normal conditions is 2.1%, and if the multiplier is 2, then a BER of 4.2% is likely. The table shows that error rates actually increase under very low stress.

**Table 3: Multipliers of "Normal" BERs by Experience, Stress, and Task Type**

|  | Very Low Stress | | |
|---|---|---|---|
| Step-by-step task | x2 | x2 | 1:1 |
| Dynamic task | x2 | x2 | 1:1 |
|  | Optimum Stress | | |
| Step-by-step task | x1 | x1 | 1:1 |
| Dynamic task | x1 | x2 | 2:1 |
|  | Moderately High Stress | | |
| Step-by-step task | x2 | x4 | 2:1 |
| Dynamic task | x5 | x10 | 2:1 |

Source: Swain and Guttman [1983], Table 18-1.

Notes:

The nominal human error probability for a task is multiplied by the modifiers. For instance, if the normal human error probability is 1.2% for a dynamic task under optimum stress done by an experienced practitioner, the estimated human error probability to use for novices under these conditions will be 1.2% x 2, or 2.4%.

Novice workers have been doing the task less than six months, while skilled workers have been doing the task more than six months. The task is not done full time during this period, simply to a normal degree.

Step-by-step tasks are guided by formal procedures. Dynamic tasks require diagnosis and decision-making, keeping track of multiple functions, and so forth.

Many believe intuitively that experienced developers make hardly any errors. Table 3 compares error multipliers for inexperienced practitioners with less than six months of experience and those with more. Beyond about six months, error rates do not go down appreciably [Swain and Guttman, 1983]. Note that only as stress increases do inexperienced versus experienced error ratios reach 2:1. It does not go higher. This limited error benefit of experience surprises many. However, an error ratio of about 2:1 for inexperienced and experienced practitioners is commonly seen in other studies comparing error rates for inexperienced and experienced people [Grudin,1993; Hayes et al., 1985; Ledgard,1980; Lesgold, et al., 1988; and Reisner 1975]

Experience increases many aspects of performance, although performance growth often plateaus for long periods or permanently [Bereiter and Scardamalia,1993]. Figure 3 shows this type of performance growth. It also shows error rates falling initially but quickly reaching a floor. Experience, even with diligent analysis and appraisal, does not decrease errors beyond some level.

**Figure 3: Experience, Performance, and Error Rates**

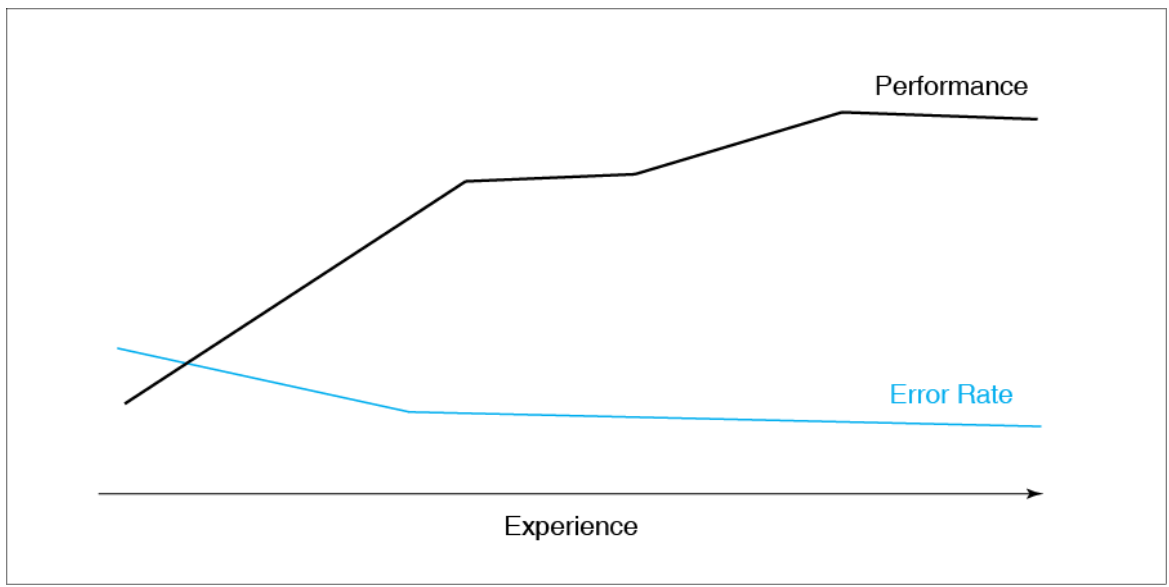

We have been looking at expertise in spreadsheet development. For spreadsheet error *detection*, in turn, Galletta et al. [1993] found that spreadsheet experience had no impact on error yield, although higher experience did decrease the time spent in inspection. Studies of error detection listed at the Panko Human Error Website often used the same spreadsheet detection task and methodology. They had approximately the same results regardless of who was doing the inspection.

# 3 PERCEIVING AND THINKING

We will begin by looking at human cognition. Humans intuitively feel that if we work carefully and check our work carefully, we will make hardly any errors. Although this is not seen in reality, it is a fundamental belief. Where does this belief come from?

Cognition involves two stages: perceiving and thinking. Perceiving encompasses our brain sensing and selecting what we will pay attention to. Thinking is how we use this information to decide what we will do. Both perceiving and thinking, we will see, are subject to extensive processing that allows us to be fast and flexible but never quite perfect. Valiant [2013] has summarized this by saying that the brain's goal is to be "probably approximately correct." This is sufficient for most situations in life, but there are situations in which error tolerances are far smaller. In these situations, we need to work very differently, and our intuition often does not see the need to do so.

## 3.1 Perceiving: Constructed Reality

Our brains do not simply pull data from our senses and draw results on a large screen in our brain for us to look at. Instead, it goes through several processing steps to develop a consistent and cohesive picture from the dizzying flood of sensations entering out bodies.

Figure 4 shows various steps that our brain uses to process the world around us based on vision, drawing primarily from Panko [2013]. Our eyes can only focus a tiny area directly in front of our gaze direction, so our gaze jerks around constantly in quick small jumps called saccades. Our brains govern these saccades, directing our gaze to things that the brain judges we need to see more clearly. It is like having a flashlight with a narrow beam and focusing it quickly in different directions. Our eyes make about 100,000 saccades per day [Mozlin, 2012]. To make the final picture coherent, however, our brains hide all of this complexity from us. We do not notice that most of what we see around us is out of focus. In addition, even if we try to pay attention to our saccades, we cannot.

**Figure 4: Constructed Reality x**

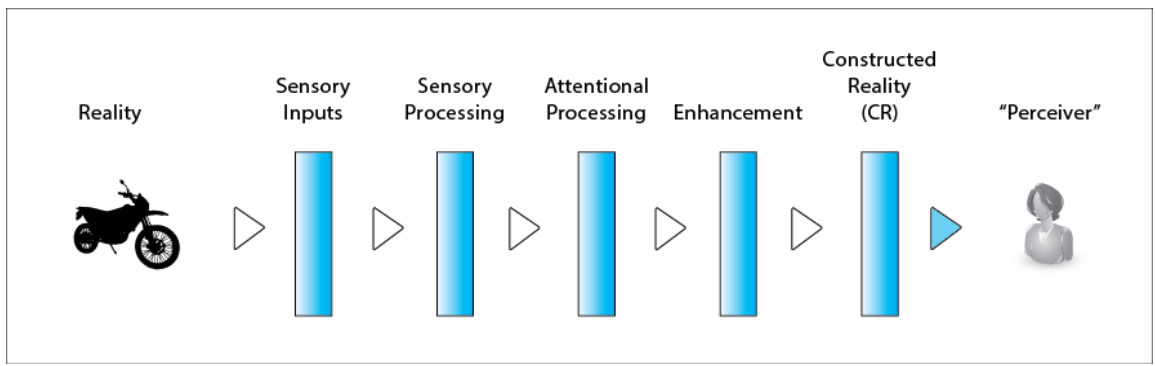

The next gate is attention. We can only pay attention to a tiny fraction of what is around us, so our brain filters out whatever it judges to be irrelevant. If we close our eyes and are asked questions about what we just saw, we will be able to recall only a few features of our environment [Chabris and Simons 2010]. However, our brain gives us the comforting illusion that we are taking in what is around us in considerable detail.

Fortunately, our brain does not merely filter out; it also enhances. For example, if we are a batter and watch a pitcher in baseball throw a ball at us, it takes our brain quite a while to get information from our eyes. To compensate, the brain adjusts for this. What we "see" in this constructed reality is the ball further along its trajectory, so that we can swing appropriately [Chow, 2013]. The brain's editing is artful editing. Again, such adjustments are hidden from us.

## 3.2 Thinking: System 1 and System 2

Constructed reality is not created for our aesthetic enjoyment. It exists to give us the ability to think about the world. It is the "data" we use in our mental computations. Unfortunately, just as constructed reality gives us a vision of the world that is probably approximately correct, while our thinking is enormously fast, it also adds to our limitations.

**System 1 Thinking**

Kahneman [2011] has noted that we appear to have two different thinking "systems"—System 1 (S1) and System 2 (S2).

System 1 is our default thinking method. Suppose that we are told that revenues are $1,000, that expenses are $800, and that we should calculate income. We will quickly say, "$200," and that is the correct answer. For obvious reasons, Kahneman [2011] calls S1 thinking "thinking fast." Thinking fast is not only fast. We are confident in its conclusions. When you thought "$200," you were not tentative in you answer.

Consider another example of thinking fast [Kahneman 2011]. Suppose that a bat costs a dollar more than a ball, and that the bat and ball together cost $1.10. How much does the ball cost? About half of us immediately give the answer ten cents. We give this answer as quickly as we did in the first task. In addition, we feel equally good about the answer.

**System 2 Thinking**

However, the correct answer is five cents. Those who realize that five cents is the correct answer probably also had the initial answer ten cents. However, they then turned on System 2 thinking. S2 thinking takes S1 answers as mere suggestions to be tested. If the ball is really ten cents, the bat must be a dollar. A dollar is not a dollar more than ten cents, of course. The correct answer is that bat is a dollar and five cents, and the ball is five cents.

In the bat and ball task, our brain did a disturbing thing. Rather than solving the actual problem, which would involve algebraic thinking, it solved a *different* problem. Kahneman [2011] notes that this frequently happens. Our brain often solves a simpler problem than the one we wish to solve. It then interpolates the answer to the actual problem. Yet when this happens, our brain gives us no indication that this has happened. More broadly, System 1 thinking gives us no indication of how well founded its conclusion is. As Kahneman has said, it does not exactly lie to us. Rather, giving us an assessment of its conclusion is not part of what it does. However, confidence in System 1 thinking cannot be used as an excuse for believing everything it suggests.

It seems like we should suspect every conclusion to be wrong. We should always use S2 thinking. There are, however, several reasons why we do not and cannot.

- First, if we actually applied S2 thinking all the time to question everything, it would drive us crazy [Gilovich 1991]. It would be an extreme form of obsessive-compulsive disorder.
- Second, we cannot sustain S2 thinking for more than a brief period. It is literally exhausting.
- Third, even when we do turn on S2 thinking, S2 thinking is difficult and certainly does not always give the correct answer. The bat and ball problem is best approached by algebra, and we all expect to make mistakes when we do algebra.

The fourth and most subtle reason we do not always use S2 thinking is that we usually do not see a need to do so. In the revenue-and-expense problem, you were probably very comfortable in your answer. As noted earlier, *whenever* S1 gives us an answer, we tend to be comfortable about it, whether it is right or wrong. We are seldom prodded to suspect the answer. Sometimes we are of course, but Allwood [1984], who observed problem solving behavior, found that only some of these instances lead to successful problem resolution. Many undetected errors generated no suspicion at all.

Allwood [1984], however, also noticed that participants sometimes simply paused to check their recent calculations. This is how they caught many of the errors that had generated no specific indications or even vague unease. This was also how they caught many of the errors that had generated unease but had not been caught earlier. Kellog [1994] found that when people write, they spend about a third of their time pausing to plan or check their work. Allwood called this activity a "standard check." We will call it unprompted error checking. Allwood noted that while this type of error checking is good it is far from being a cure-all for errors.

## 4 IMPROVING SPREADSHEET ACCURACY

In the first section of this paper, we saw that there are substantial, compelling, and unanimous research findings that spreadsheets are at high risk of spreadsheet errors. In the next section, we saw how human cognition, in both perception and thinking, give us an approximate view of the world that works most of the time but that hides information from us, giving us a distorted experience of what errors we make and how good we are at finding errors.

If the research is true, and if we wish to act on empirical findings like a pilot trusting his or her instruments in darkness instead of physical sensations about an airplane's orientation in the sky, what should we do? Figure 5 may give us some guidance. To create a program or spreadsheet, the creator usually begins with modules that are later integrated into larger units and finally into whole programs or spreadsheets. In software development, module testing is called unit testing. Spreadsheet development experiments create spreadsheet models that are similar in size to software modules. The figure shows that error rates are similar at the end of the module stage (but before extensive unit testing) for spreadsheets and programs.

Figure 5: Error Rates at the Module and Final Stages of Development

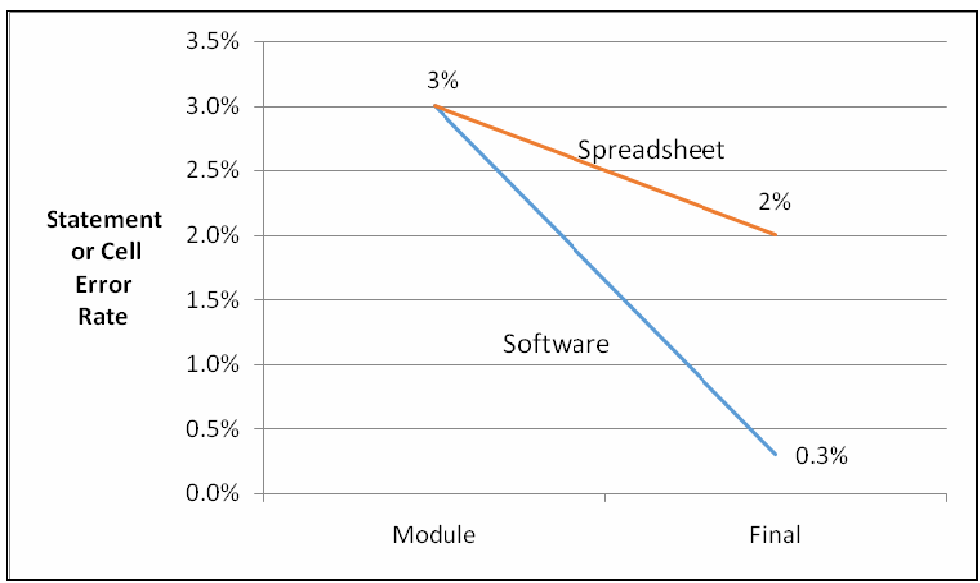

We saw this similarity earlier. The figure also shows final error rates in delivered programs, based on errors found after delivery. Note that roughly 90% of the errors in a program are removed during development.

We saw earlier that commercial spreadsheet inspections (audits) and testing of similar intensity found cell error rates of 1.2%, 2.2%, and 2.5%. These inspections averaged about one person-week apiece. Powell, Baker, and Lawson [2008] only found a cell error rate of 0.9% when hardcoding, which all other studies treat as a qualitative error, was removed. However, the average time spent per spreadsheet was only 3.25 hours. Clermont, Hanin, and Mittermeier [2000] found a quantitative cell error rate of only 0.4%. However, it only used static analysis auditing software. Anderson [2004] and Aurigemma and Panko [14] compared errors discovered with this type of software to actual errors in spreadsheets and found that they were not effective in finding errors. Anderson's study found that the programs he used only caught 27% of the errors he seeded in the spreadsheet used in the experiment. Aurigemma and Panko's seven testers found far fewer errors made by actual developers (natural errors instead of seeded errors). Aurigemma and Panko, however, did not use the mapping functions available in this software, although the spreadsheets did not have a strong block structure.

Of course, while cell error rates (CERs) at the module and final stage are similar, they should not be. Inspections of final spreadsheets should have undergone such extensive testing that few errors should remain—as in the case of software development. Furthermore, while the spreadsheets developed in experiments have known correct solutions, all errors can be found. In turn, inspections of operational spreadsheets only report CERs based on discovered errors. Given human limitations in discovering errors, the inspection CERs are likely to be larger than the three numbers shown. Unfortunately, none of the inspections of operational spreadsheets had multiple inspectors inspect the spreadsheet individually and report their errors. This would have allowed a statistical estimate of remaining errors, so that error detection effectiveness could be assessed.

What is different between software development and spreadsheet development? Initially, programmers attempted to reduce errors by writing structured programs and using languages with such features as strong data typing. While this helped somewhat, the results of improved development practices were modest. Brooks [1987] summarized this modest progress succinctly. He said that there were no silver bullets in software development.

In the end, commercial software developers addressed the problem of errors by focusing primarily on testing. They accepted that errors would occur and would continue to occur in acceptable numbers even when good development practices are used. Over time, they had to keep increasing testing until they reached acceptable final error rates. Jones (1998) looked at 84 projects in 27 organizations. Testing time ranged from 27% to 34% of total development resources. In every case, subjects reported that not enough time was spent testing. Later, Kimberland (2004) reported that Microsoft software development teams spent 40% to 60% of their total working time in testing.

For software, dynamic testing based on well-chosen test cases with oracles (correct answers) is the most common form of testing. This is good when there is simple final behavior. For instance, if the software is designed to copy changed authentication credentials to all authentication servers, it is relatively easy to see if the software does this under various plausible scenarios. In spreadsheets, however, results are derived from chains of calculations. For each calculation, test cases for a dozen or more equivalence classes may have to be undertaken. This would result in a combinatorial explosion in test cases to compute a final value. Furthermore, spreadsheet software does not have native functionality for implementing this.

For the unit testing of new or modified modules, commercial software companies often turn to static analysis, in which the code is examined without running it. There are several informal methods for static inspection, but firms that use static analysis typically use a specific formal method called code inspection. The code inspection methodology was originally developed by Fagan [1976, 1986] and was later developed by others. Today, it is normal to replace the final meeting in which errors are reported, assessed, and accepted or not accepted with an online process to do these things [Cohen, 2006]

Fagan [1976, 1986] grounded code inspection in empirical analysis. A prime tenet was the obligation to report on the broad results of code inspections, at least in the aggregate. Consequently, code inspection has produced data on tens of thousands of commercial code inspections. This resulted in refinements as different approaches were actually proven to produce better results. People who argued that things should be done in a certain way because they were experts often found themselves wrong. Good practice emerged based on merit, not on the authority of the proposer.

Although Fagan [1976 1986] and later developers did not base their work on human error research, the code inspection is deeply consistent with this research.

A code inspection is always done by teams, never by individuals. Teams typically have three to eight members. This is consistent with the fact that humans are only modestly effective at finding errors. Even with team inspection, code inspections typically only find 60% to 80% of all errors in a module[Boehm and Basili 2001, Eick, et al. 1992, Fagan 1986, Hall 1996, Jones 1986, Jones 1998]. The rest must be found in later testing. Based on data provided from code inspections, single-person code inspection would only catch 20% to 40% of all errors. Panko [1999] conducted an experiment applying code inspection methodology to spreadsheet inspection. He found that individuals only caught 63% of all errors in the spreadsheet they examined. In teams of three, the average was 83%. This is a modest percentage increase, but produced large jumps in detection rates for the most difficult-to-find errors. Inspection by individuals would be difficult to justify in spreadsheet developers.

Code inspection always begins with an initial meeting in which the ground rules are laid out, roles are assigned, and, most importantly, the team is taken through the program line by line to ensure that all inspectors understand the purpose of the module, the purpose of each part, and how the code works. During this process, some errors may be noted by the inspectors, but that is not the purpose of the initial meeting. The initial meeting is important because program code is not self-documenting, even when comments are used. (As developers sometimes remark, "That's why they call it code."). In their in-depth interviews with spreadsheet developers, both Nardi and Miller [1991] and Hendry and Green [1994] noted that this problem is serious with spreadsheets. While the rectangular structure of worksheets is ideal for some models, it sometimes creates difficult challenges. Difficult challenges tend to lead to complex precedent structures. Joseph [2000] conducted a code inspection experiment on spreadsheets. He confirmed the importance of the initial meeting, especially for complex logic flows.

After the initial meeting, the inspectors work on the code individually. Initially, the only goal was to develop a very good understanding of the code. Over time, it became more common to make this the stage for doing most error discovery and recording. This stage also has been the most studied and analyzed empirically.

Due to limits of human vigilance, individual inspections must be limited to about an hour. Inspection is mentally exhausting, and longer inspection periods produce a sharp drop in error identification. The statement inspection rate also must be limited to about 100 lines of code per hour for best results. Faster inspection rates again give lower error detection yield. Taken together, these limitations mean that code inspection can only be applied to software modules, not entire programs. Give the data on how rapidly error yield decreases over longer periods and correct inspection rates, and given the importance of understanding what one is inspecting, inspecting an entire program makes no sense.

In the review meeting, the team meets again. Initially, this was the stage in which errors were sought by active inspection. It gradually became the time to gather reports, determine if they are legitimate errors, rate the severity of errors, and create a list of errors to be removed. The team does not attempt to recommend fixes for these errors. Unfortunately, the need to have a review meeting often delayed the results of code inspections. Today, this stage is done online using application software designed for this work. In fact, the software now manages all stages.

After the review meeting, the code inspection is documented. Based on this documentation and data collected by the application, patterns for meetings can assessed. In addition, there needs to be error tracking to verify that errors are corrected.

In spreadsheet development, in contrast, all studies since the 1980s have found that detailed inspection is rare. When it does exist, it typically uses single inspectors and covers chunks of code that are very large with fast inspection rates.

We argue, based on human error research and specific software and spreadsheet research, that code inspection should be used to test spreadsheets. Empirical analysis will be needed to develop size and speed metrics for spreadsheet inspection, but we are likely to find that we will need to spend 20% to 40% of our development resources on testing. Doing this will be expensive. Not doing it is likely to be unprofessional.

## 5. CONCLUSION

In this paper, we saw that there is substantial, compelling, and unanimous evidence that spreadsheet errors are occurring at a rate that corporations should see as unacceptable. Humans have many intuitions about errors, but the way our cognition works means that we are conscious of very few of our errors. Consequently, we tend to believe, among other things, that few spreadsheets have material errors, that we make few errors, that anyone can make only rare errors and catch almost all errors if they just tried harder and spent more time, and that we know how to reduce spreadsheet errors through good practice.

If we accept that our "experiences" regarding errors is impaired, so that we must turn to human error research, software industry experience, and spreadsheet error research, then we need to do far more testing. Testing is not merely one of many error controls. It is likely to be by far the most important error control. When asked if their spreadsheet contains an error, developers in one unpublished study often said something to the effect that they did not see how that would be possible because they spent "so much time" testing it—although they always added that, of course, everyone makes mistakes. When asked if they were spending 20% to 40% of their time doing testing, their typical response was something to the effect that that would be crazy. If spreadsheet development is to be regarded as a true profession, the biggest change is likely to be a practice requirement for very extensive and expensive testing.

What about the many prescriptions that are touted for reducing spreadsheet errors? Many begin with a statement that, "the problem with spreadsheets is X." X may be having end users do development instead of professionals, a failure to use a program with strong data typing, or many other things. Although proposers know that they are correct because of their "experience," this knowledge should not be viewed as reliable. In addition, almost all prescriptions are for development practices, and if there are testing prescriptions as part of these practices, they are usually rudimentary. Senders and Moray [1991], in their review of human error research, noted that many error reduction prescriptions have bad side effects. For example, requiring strong typing could put more burden on developers who are already overloaded and making errors as a result. Given limits in human attention and other resources, adding burdens could increase errors much more than it reduces the type of error it was intended to reduce. Given that out experience of errors in unreliable, intuitions about how to reduce errors should not be taken seriously unless they are rigorously tested. As in the case of code inspection, we can draw from human error literature and proven practices in other areas and test whether it is feasible and effective to apply them to spreadsheets. Medicines must be tested to prove that they are safe and effective before they are accepted. We should require the same rigor for spreadsheet development prescriptions.

In software development, many students and professionals have adopted Watts Humphrey's Personal Software Process (PSP), in which the person keeps detailed information on their software development activities. The goal of PSP is to counteract intuitive ideas about software errors with actual data. The results are typically a shock to the programmer using it. However, the goal is not to shock but to help the programmer develop better estimation skills for time, resources, and the work needed to reduce errors.

When people are confronted with their actual error rates they are typically shocked. For example, one study examined how frequently people hit the wrong pedal when they drive. The project director was worried that the practice would be so rare that measuring it would be extremely difficult. When the team did the study, they found to their amazement that the frequency of hitting the wrong pedal was about once per hour [Spiegel 2010]. Self-knowledge is the most potent force for navigating through work.

**REFERENCES**

Allwood, C. M. (1984) "Error Detection Processes in Statistical Problem Solving," *Cognitive Science,* (8:4) October-December, pp. 413-437.

Anderson, J. (2007). "Study: Electronic Record-Keeping Doesn't Improve Health Care," ergoweb.com. http://www.ergoweb.com/news/detail.cfm?id=2127. Last viewed October 12, 2014.

Aurigemma, S. A. and Panko, R. R. (2014)."Evaluating the Effectiveness of Static Analysis Programs versus Manual Inspection in the Detection of Natural Spreadsheet Errors," *Journal of Organizational and End User Computing*, (26,1) January 2014, 47-65.

Basili, V. R., & Selby, R. W., Jr. (1986). Four Applications of a Software Data Collection and Analysis Methodology. In J. K. Skwirzynski (Ed.), *Software System Design Methdology* (pp. 3-33). Berlin: Springer-Verlag.

Bereiter, C., and Scardamalia, M. (1993). *Surpassing ourselves: An inquiry into the nature and implications of expertise*. Chicago: Open Court.

Boehm, B. & Basili, V. R. (2001, January). "Software Defect Reduction Top 10." *Computer*, 135-137.

Bridgeman, B., Hendry, D., and Stark, L. (1975). "Failure to Detect Displacement of the Visual World during Saccadic Eye Movements,"*Vision Research*(15),pp. 719-722.

Brooks, Fred P. (1987). "No Silver Bullet—Essence and Accidents of Software Engineering". *IEEE Computer*, 20(4): 10–19.

Chabris, C., and Simons, D. (2010). *The Invisible Gorilla: And Other Ways our Intuitions Deceive Us,* New York: Crown.

Chow, D. 2013, How Your Brain Tracks Moving Objects, May 8. http://www.livescience.com/29417-how-brain-tracks-moving-objects.html.

Cohen, J. (2006). *Best kept secrets of peer code review*. Austin Texas: Smart Bear.

Eick, S. G.; Loader, C. R.; Long, M. D.; Votta, L. G.; & Vander Weil, S. (1992). "Estimating Software Design Faults," *Proceedings of the Fourteenth International Conference on Software Design,* Melbourne, May, pp. 59-65.

Fagan, M.E. (1976). Design and code inspections to reduce errors in program development. *IBM Systems Journa*l, 15(3), 182-211.

Fagan, M.E. (1986). Advances in software inspections. *IEEE Transactions on Software Engineering,* SE-12(7), , July, 744-751.

Galletta, D. F., Abraham, D., El Louadi, M., Lekse, W., Pollailis, Y. A., and Sampler, J. L. (1993). An Empirical Study of Spreadsheet Error Performance. Journal of Accounting, Management, and Information Technology, 3(2), 79-95.

Gilb, T. and Graham, D. (1993). Software inspection. Edinburgh Gate, England: Person Education.

Gilovich, T. (1991). *How We Know What Isn't So: The Fallibility of Human Reason in Everyday Life,* New York: Free Press.

Graden, M., Horsley, P., and Pingel, T. (1986). The effects of software inspection on a major telecommunications project. *ATandT Technical Journal*, 65.

Grudin, J. (1983). Error patterns in skilled and novice transcription typing. In W. E. Cooper (Ed.), *Cognitive aspects of skilled typewriting* (pp. 121-143). New York: Springer-Verlag.

Hall, A. (1996). Using Formal Methods to Develop an ATC Information System. IEEE Software, 13(2), 66-76.

Hayes, J.R., Flower, L.S., Schrivner, K.J.S., and Carey, L. (1985). *Cognitive processes in revision* (Technical Report No. 12). Pittsburgh: Carnegie Mellon University, Communications Design Center.

Hendry, D.G., and Green, T.R.G. (1994). Creating, comprehending, and explaining spreadsheets: A cognitive interpretation of what discretionary users think of the spreadsheet model. *International Journal of Human–Computer Studies*. 40(6), 1033-1065.

Hermans, F., and Murphy-Hill, E. (2015). "Enron's Spreadsheets and Related Emails: A Dataset and Analysis" 37th International Conference on Software Engineering, (ICSE).

Hicks, L. (1995). NYNEX, personal communication with the first author via electronic mail.

Johnson, P., & Tjahjono, D. (1997). *Exploring the Effectiveness of Formal Technical Review Factors with CSRS.* Proceedings of the 1997 International Conference on Software Engineering, Boston, MA, May.

Jones, C. (1986). *Programming Productivity*, McGraw-Hill, New York.

Jones, T. C. (1998). *Estimating Software Costs*, New York" McGraw-Hill.

Joseph, Jimmie L. (2000). The Effect of Group Size on Spreadsheet Error Debugging, PhD Dissertation, University of Pittsburgh.

Kahneman, D. (2011). *Thinking, fast and slow*, New York: Farrar, Strauss and Giroux.

Kellog, R.T. (1994). *The Psychology of Writing,* New York: Oxford University Press.

KPMG Management Consulting (1998). Supporting the decision maker: A guide to the value of business modeling. press release, July 30.

Ledgard, H.F., Singer, A., Whiteside, J.A., and Seymour, W. (1980). Directions in human factors for interactive systems. *Communications of the ACM,* 23(10), 556-563.

Lesgold, A., Rubinson, H., Feltovich, P., Glaser, R., Klopfer, D., and Wang, Y. (1988). Expertise in a complex skill: diagnosing X-Ray Pictures. In R.G.M. Chi, and M. Farr (Eds.), *The nature of expertise*. Hillsdale, NJ: Lawrence Erlbaum.

Lukasik, T., CPS. (1998). Personal communication with the author, August 10.

McDaid, K., MacRuairi, R., Clynch, N., Logue, K., Clancy, C., and Hayes, S. (2011). Spreadsheets in financial departments: An automated analysis of 60,000 spreadsheets using the Luminous Map technology. *Proceedings of the 2011 European Spreadsheet Risks Information Group*.

Mozlin, R. (2012). *100,000 saccades per day*. College of Optometrists in Vision Development Blog, January 6. http://covdblog.wordpress.com/2012/01/06/100000-saccades-per-day/.Last viewed June 21, 2015.

Myers, G. J. (1978). A Controlled Experiment in Program Testing and Code Walkthroughs/Inspections. *Communications of the ACM, 21*(9), 760-768.

Nardi, B. A., and Miller, J. R. (1991). Twinkling Lights and Nested Loops: Distributed Problem Solving and Spreadsheet Development. International Journal of Man-Machine Studies, 34(1), 161-168.

O'Neill, D. (1994). National Software Quality Experiment. *4th International Conference on Software Quality Proceedings*, October

Panko, R. R. (2015a). Human error website, http://panko.com/HumanErr.Last viewed June 21, 2015.

Panko, R. R. (2015b). Spreadsheet error website. http://panko.com/ssr.Last viewed June 21, 2015.

Panko, R.R. (2013). The cognitive science of spreadsheet errors: Why thinking is bad. *Proceedings of the 46th Hawaii International Conference on System Sciences*, Maui, Hawaii: IEEE, January 7-11.

Panko, R. R. (1999). "Applying Code Inspection to Spreadsheet Testing," *Journal of Management Information Systems*, 16(2), Fall, 159-176.

Panko, R.R., and Halverson, R.P., Jr. (2001). An experiment in collaborative spreadsheet development. *Journal of the Association for Information Systems* 2(4), July.

Panko, R. R. and Halverson, R. H. Jr. (1996). "Understanding Spreadsheet Risks," *Office Systems Research Journal*,14(2), Fall 1-11.

Panko, R.R. and Sprague, R. H., Jr. (1998). "Hitting the Wall: Errors in Developing and Code Inspecting a 'Simple' Spreadsheet Model," *Decision Support Systems*, 22(4), 337-353.

Porter, A., Votta, L. G., Jr., & Basili, V. R. (1995). Comparing Detection Methods for Software Requirements Inspections: A Replicated Experiment. *IEEE Transactions on Software Engineering, 21*(6), 563-575.

Porter, A. A., & Johnson, P. M. (1997). Assessing Software Review Meetings: Results of a Comparative Analysis of Two Experimental Studies. *IEEE Transactions on Software Engineering, 23*(3), 129-145.

Porter, A. A., Sly, H. P., Toman, C. A., & Votta, L. G. (1997). An Experiment to Assess the Cost-Benefits of Code Inspections in Large Scale Software Development. *IEEE Transactions on Software Engineering, 23*(6), 329-346.

Porter, A. A., & Votta, L. G. (1994, May 16-21, May 16-21). *An Experiment to Assess Different Defect Detection Methods for Software Requirements Inspections*. Proceedings of the 16th International Conference on Software Engineering, Sorento, Italy.

Reason, J.T. (1990). *Human error*. Cambridge, UK: Cambridge University Press.

Reisner, P., Boyce, R.F., and Chamberlin, D.D. (1975). "Human factors evaluation of two data base query languages—SQUARE and SEQUEL,"*Proceedings of the National Computer Conference, Arlington, Virginia*.

Senders J.W. and Moray, N.P. (1991). *Human Error: Cause, Prediction, and Red*uction, LawrenceErlbaum, Hillsdale, NH.

Spiegel, A. (2010). "Runaway Cars: Driver Error or Car Malfunction?" National Public Radio, March 19, http://www.npr.org/2010/03/19/124815144/runaway-cars-driver-error-or-car-malfunction. Last viewed June 21, 2015.

Swain, A.D., and Guttman, H. E. (1983). *Handbook of human reliability analysis with emphasis on nuclear power plant applications* (Technical Report NUREG/CR-1278). U. S. Nuclear Regulatory Commission, Washington, DC.

Valiant,L.G. 2013.*Probably Approximately Correct*, New York: Basic Books.

Weller, M. (1993). Lessons from three years of inspection data. *IEEE Software*, 10(5), 38-45.

Woods, D. D. (1984). "Some Results on Operator Performance in Emergency Events," *Institute of Chemical Engineers Symposium Series, 90,* pp. 21-31. Cited in Reason, 1990.